\documentclass[reprint, amsmath, amssymb, aps, superscriptaddress, prb, longbibliography]{revtex4-2}

\usepackage{color}
\usepackage{graphicx}
\usepackage[colorlinks=true, urlcolor=blue, linkcolor=blue, citecolor=blue]{hyperref}
\usepackage{ulem}
\usepackage{footnote}

\usepackage{xcolor}

\begin{document}
	
\title{
Transport properties of a 1000-nm HgTe film:\\
the interplay of surface and bulk carriers
}

\author{M.\,L.\,Savchenko}\email{mlsavchenko@isp.nsc.ru}
\affiliation{Institute of Solid State Physics, Vienna University of
	Technology, 1040 Vienna, Austria}
\affiliation{Rzhanov Institute of Semiconductor Physics, 630090 Novosibirsk,
	Russia}

\author{D.\,A.\,Kozlov}
\affiliation{Experimental and Applied Physics, University of Regensburg, D-93040 Regensburg, Germany}
\affiliation{Rzhanov Institute of Semiconductor Physics, 630090 Novosibirsk,
	Russia}

\author{N.\,N.\,Mikhailov}
\affiliation{Rzhanov Institute of Semiconductor Physics, 630090 Novosibirsk,
	Russia}
\affiliation{Novosibirsk State University, 630090 Novosibirsk, Russia}

\author{S.\,A.\,Dvoretsky}
\affiliation{Rzhanov Institute of Semiconductor Physics, 630090 Novosibirsk,
	Russia}

\author{Z.\,D.\,Kvon}
\affiliation{Rzhanov Institute of Semiconductor Physics, 630090 Novosibirsk,
	Russia}
\affiliation{Novosibirsk State University, 630090 Novosibirsk, Russia}

\begin{abstract}
\noindent
We report on systematic study of transport properties of a 1000-nm HgTe film. 
Unlike to thinner and strained HgTe films, which are known as high-quality three-dimensional (3D) topological insulators, the film under study is much thicker than the limit of pseudomorphic growth of HgTe on a CdTe substrate. 
Therefore, it is expected to be fully relaxed and has the band structure of bulk HgTe, i.e., a zero gap semiconductor. 
Nevertheless, since the bands inversion the two-dimensional (2D) topological surface states are still expected to exist. 
To check this claim we studied classical and quantum transport response of the system.
We demonstrate that by tuning the top-gate voltage one can change the electron-dominating transport to the hole one.
The highest electron mobility
is found to be more than $300 \times 10^3$\,cm$^2$/Vs. 
The system exhibits Shubnikov-de Haas (SdH) oscillations with a complicated pattern and shows up to 5 independent frequencies in corresponding Fourier spectra. 
They are attributed to the topological surface states, Volkov-Pankratov states and spin-degenerate bulk states in the accumulation layer near the gate. 
The observed peculiarities of the quantum transport are the strong SdH oscillations of the Hall resistance, and the suppressed oscillatory response of the topological surface states. 
\end{abstract}
	
	\date{\today}
	\maketitle

\section{Introduction}
HgTe and HgCdTe crystals and films have been intensively studied for more than 50 years~\cite{Berchenko1976, Chu2008}.
At the first, the most intriguing phenomenon in these systems was a zero or close to zero bulk band gap.
It results in the sharp increase of spin-orbit corrections to the Hamiltonian that modifies the dispersion of the system, and makes it possible to study ultrarelativistic particles in the solid state and check the theoretical predictions related their properties. 
Despite difficulties relating the pour quality and instability of the first HgTe devices, the most important characteristic of a HgTe semiconductor -- its dispersion -- was obtained, but only for bulk carriers of the system.
It was successfully found how the dispersion depends on structure composition and temperature.
Moreover, it has been predicted~\cite{Dyakonov1981a, Volkov1985} and accidentally found~\cite{Minkov1997} that HgTe 
has peculiar non-degenerate surface states 
that now called ``topological surface states''.

Following the strong increase of the crystal quality of HgCdTe systems and clarifying surface states properties, especially spin-momentum locking and linear dispersion, there was reopening of rich physics in HgTe-based structures \cite{Hasan2010, Ando2013}. 
Now it is well-known that there are topologically non-trivial surface states in strained 2D or quasi-2D HgTe films, which exist at all accessible in the experiment Fermi level positions regardless of the bulk band gap existence~\cite{Brune2011, Kozlov2014, Savchenko2019}.

However, apart from magneto-optic measurements~\cite{Shuvaev2011, Shuvaev2012, Shuvaev2013, Hubmann2020, Otteneder2020} and a short transport report~\cite{Brune2011}, there was no systematic study of transport properties of bulk HgTe, where topological surface states are 
taking into account. Compare to previously studied 80- and 200-nm HgTe films, a much thicker 1000-nm film has trivial bulk 3D carriers that can interact with topological surface states and modify their transport response. 
Moreover, a high spatial separation of the surface states results in their full electrostatic decoupling, making this object promising to study only one spin-degenerate topological surface. 

In this paper, we report the study of transport properties of a 1000-nm HgTe film.
The analysis and comparison of classical and quantum magnetotransport allows us to identify several groups of carriers.
According to our results, the system can be tuned from a low-mobility mixed electron and hole bulk transport regime to the 2D mode, when high-mobility electrons or holes located in the accumulation layer near the gate dominate the transport and exhibit pronounced SdH oscillations.

\section{Methods}
\begin{figure}
	\includegraphics[width=1\columnwidth]{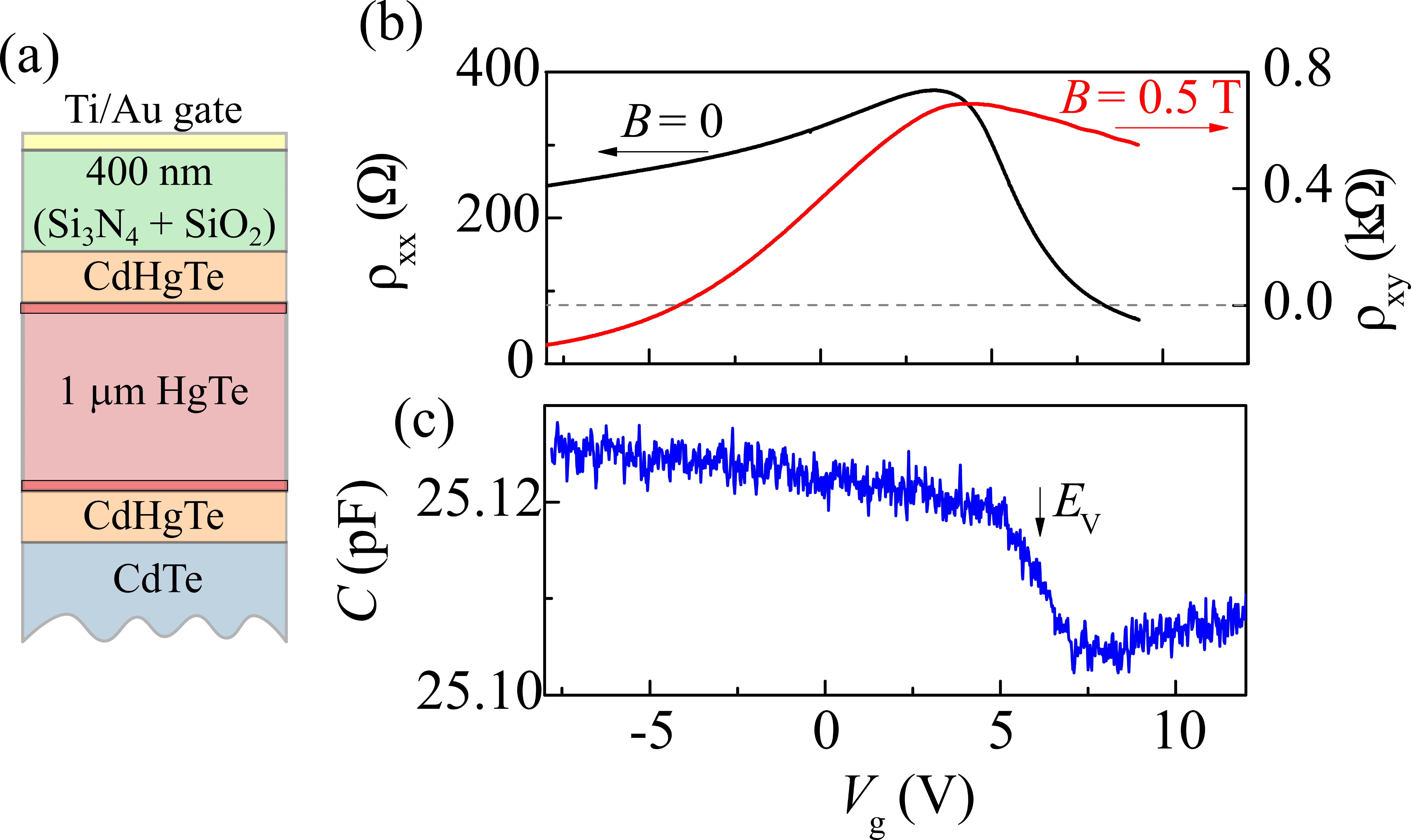}
	\caption{
		(a)~Schematic cross-section of the structure under study. 
		The 1000-nm~HgTe film is placed between thin Cd$_{0.6}$Hg$_{0.4}$Te barrier  layers covered by a 200+200\,nm Si$_3$N$_4+$SiO$_2$ insulator and a metallic gate.
		Bright red lines represent the surface states on the top and bottom surfaces of the HgTe film.
		(b)~The gate voltage dependences of longitudinal resistance $\rho_\text{xx}$ (black line, left axis, zero magnetic field) and Hall resistance $\rho_\text{xy}$ (red line, right axis, magnetic field $B = 0.5\,$T).
		(c)~The gate voltage dependences of capacitance $C$ measured at frequency 222\,Hz.
		The arrow indicates the saturation of a sharp increase of $C$ that corresponds to the Fermi level position near the valence band top.
	} \label{Fig1}
\end{figure}

Measurements are carried out on 1000-nm HgTe films that have been grown by molecular beam epitaxy on a GaAs(013) substrate at the same conditions and with the same layer ordering as it was for usual 80 and 200\,nm films~\cite{Kozlov2014, Savchenko2019}.
But apart from previously studied systems, the studied structure is a clear 3D system in terms of the bulk sub-bands formation.
In Fig.~\ref{Fig1}\,(a) we schematically show a cross-section view of the system under study.
The HgTe film is placed between thin Cd$_{0.6}$Hg$_{0.4}$Te buffer layers, a Ti/Au gate has been deposited on the 200+200\,nm Si$_3$N$_4+$SiO$_2$ insulator grown by a low temperature chemical vapor deposition process.
An approximate thickness of the pseudomorphic growth of a HgTe film on a CdTe substrate with a 0.3\%
larger lattice constant is about 100 -- 150\,nm  (according to~\cite{Brune2011} and our experience).
Thereby we believe that our 10 times thicker HgTe film is fully relaxed to its own lattice constant and is a zero gap semiconductor 
\cite{Berchenko1976}. 

The studied Hall-bars have a 50\,$\mu$m current channel and equal to 100 and 250\,$\mu$m distances between potential probes. 
Transport measurements were performed using a standard lock-in technique with a driving current in the range of 10$^{-10}$ -- 10$^{-7}$\,A in a perpendicular magnetic field $B$ at temperature 0.2\,K. 
A current frequency for transport measurements was 12\,Hz. 
For the capacitance measurements we mixed the dc bias $V_\text{g}$ with a small ac voltage $V_\text{ac}$ and measured the ac current flowing across the device phase sensitively. 
The total capacitance measured in such a way between the metallic top gate and a two-dimensional electron-hole system depends, besides the geometric capacitance, on the quantum capacitance $e^2D$, connected in series and reflecting the finite density of states $D$ of the system~\cite{Smith1985, Kozlov2016}; $e$ is the elementary charge, $D$ is the thermodynamic density of states.
The ac frequency for capacitance measurements was in the range of 0.2 -- 3\,kHz.
The frequency independence of measured resistance and capacitance $C$ was controlled excluding both the existence of leakage currents and resistive effects. 
The parasitic capacitance of our set up is about 20\,pF.

\section{Results and discussion}\label{Sec: Results and discussion}

The system under study can have several groups of carriers. 
There are trivial 3D bulk electrons and holes.
An introduced by the gate voltage accumulation layer can host either electrons or holes that have 2D nature.
Besides them the system is expected to have the 2D topological surface states at all gate voltages~\cite{Volkov1985}.
These states are non-degenerate and located near the top (closer to the gate) and bottom (closer to the substrate) surfaces.
Additionally, one may expect to detect the response from  spin-degenerate  Volkov-Pankratov states~\cite{Volkov1985}.
They have the same origin as topological surface states~\cite{Inhofer2017, Tchoumakov2017}, however they form only if there is a smooth transition between topologically trivial and non-trivial materials, 
or if there is strong enough band bending near the boundary of materials with the opposite topology~\cite{Tchoumakov2017}.
Note that the high density of states of 3D carriers pins the Fermi level in the bulk. Therefore, the gate voltage changes the charge state of the system primarily near the gate.

\subsection{Classical transport and Drude fitting}\label{SubSec: Classic transport}

In Fig.~\ref{Fig1}\,(b) we show the examples of the gate voltage dependences of the longitudinal resistance $\rho_\text{xx}$ (black line, left axis, zero magnetic field) and the Hall resistance $\rho_\text{xy}$ (red line, right axes, magnetic field  $B = 0.5\,$T). 
The gate voltage dependence of capacitance $C$ is presented in panel (c).
A similar picture was observed earlier on thinner HgTe films~\cite{Kozlov2014, Savchenko2019} and has the following explanation. 
The measured capacitance is represented as two capacitors connected in series. The first capacitor reflects the geometric capacitance, whose value is determined by the distance from the gate to the center of the carrier wave function. The second capacitor represents quantum capacitance and its value is proportional to the density of states. In our system, due to screening effects, the measured capacitance is sensitive to the density of states of the carriers closest to the gate, i.e., those in the accumulation layer or topological surface electrons. Carriers located in the bulk practically do not influence the measured capacitance. 
At large positive gate voltages the system exhibits electron-dominated transport.
Moving $V_\text{g}$ to its lower values we decrease the electron density, increase the longitudinal and Hall resistance, while the capacitance goes down because of a small increase in the distance from electrons to the gate.
A sharp increase of capacitance in the region from 7.5 to 5\,V, indicates the Fermi level enters the valence band for the carriers, located near the gate.
The capacitance increase governs by about ten times higher effective mass and consequently density of states of holes compare to electrons in HgTe~\cite{Gospodaric2019}. However, the sign of the Hall resistance still shows the electron-dominated transport, indicating the co-existence of electrons and holes in this region of the gate voltages.
A further gate voltage decrease results in both an electron density decrease and a hole density increase.
Since the electron mobility is higher compare to the hole one, the resistance maximum $\rho_\text{xx}^\text{max}$ is to the left from $E_\text{v}$ at about $V_\text{g}^\text{max} \approx 4\,$V. 
At about $-3$\,V the Hall resistance changes its sign, confirming the transition to the hole-dominated transport, though the electrons are still present in the system (see below).
Thus, depending on $V_\text{g}$, holes or electrons dominate a transport response.

\begin{figure}
	\includegraphics[width=1\columnwidth]{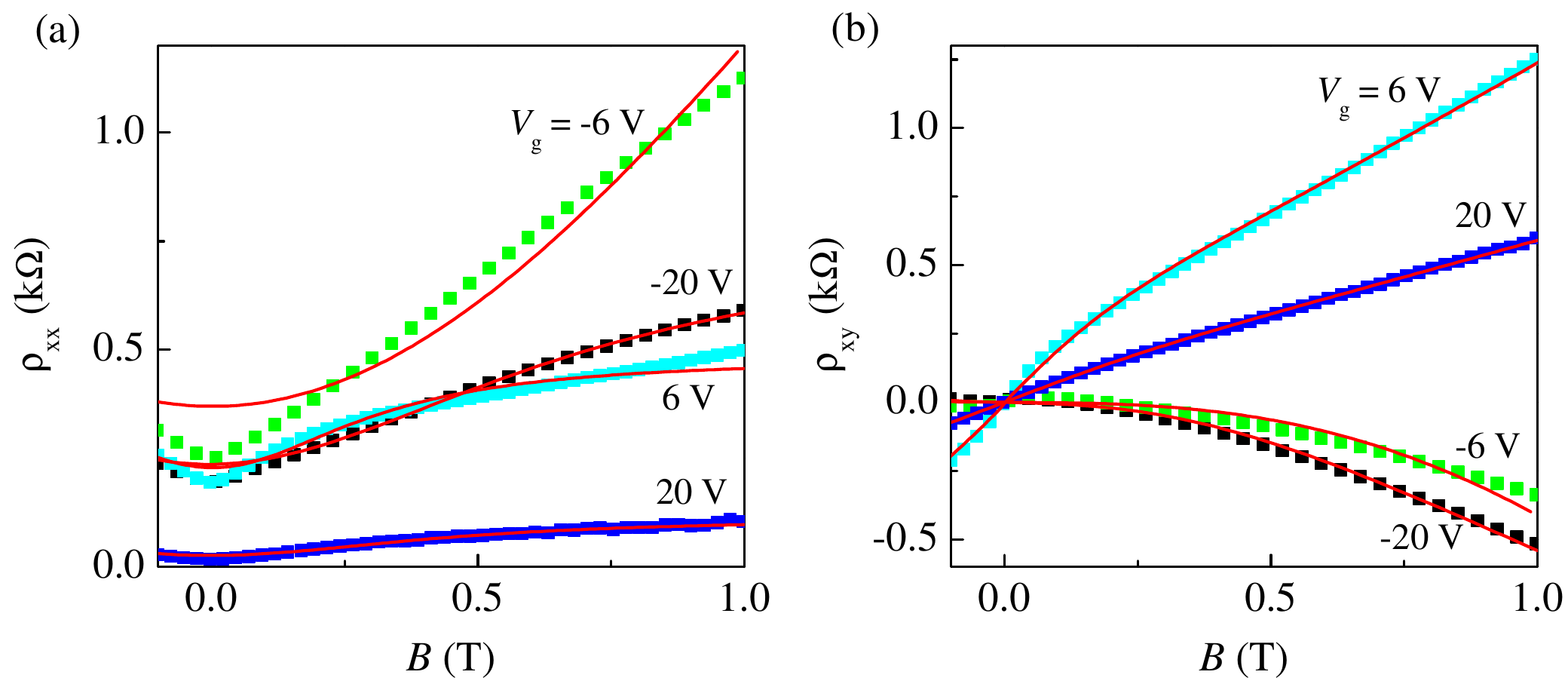}
	\caption{
		The examples of magnetic field dependences of $\rho_\text{xx}$~(a) and $\rho_\text{xy}$~(b) measured at different gate voltages. 
		Red solid lines represent the two-component Drude fittings.
	} \label{Fig2}
\end{figure}

In Fig.~\ref{Fig2}\,(a) and (b) we show examples of magnetic field dependences of longitudinal $\rho_\text{xx}$ and Hall resistance $\rho_\text{xy}$ measured at $V_\text{g} = -20,\,-6,\,6$ and 20\,V. 
There is strong positive magnetoresistance in $\rho_\text{xx}(B)$ at all gate voltages~(see Supplementary Fig.~S2) that together with nonlinear $\rho_\text{xy}(B)$ indicate the co-existence of several types of carriers in the system.
At high magnetic fields $\rho_\text{xy}$ is determined by the total  charge carrier density, resulting in a different sign of the Hall resistance for high positive gate voltages (where there are only electrons) and negative gate voltages (where holes dominate transport).
On the contrary, in weak magnetic fields and at all gate voltages a positive or near-zero slope of $\rho_\text{xy}(B)$ is observed~(see Supplementary Fig.~S2). 
According to the multi-component Drude model ~\cite{Kozlov2014, Savchenko2016, Vries2017, Ziegler2020}, in weak magnetic fields the slope of the Hall resistance is determined, to a large extent, by carriers with high mobility and indicates the presence of high mobility electrons (at all gate voltages) on a background of electrons or holes with much lower mobility.
Within the model, the conductivity tensor components are equal to the sum of partial conductivities: the diagonal component of the conductivity tensor is equal to $\sigma_\text{xx} (B) = \sum e n_\text{i} \mu_\text{i}/\left(1+(\mu_\text{i}B)^2\right)$, 
and the Hall conductivity is equal to $\sigma_\text{xy} (B) = \sum \text{sign}(i) e n_\text{i} \mu_\text{i}^2B/\left(1+(\mu_\text{i}B)^2\right)$, where $n_\text{i}$ and $\mu_\text{i}$ are density and mobility of the carriers and $\text{sign}(i)$ denotes the sign of the carriers in the Hall signal.

The Drude model works for any number of groups of carriers, however the model tolerance is too high to distinguish more than two groups reliably. 
Therefore, we fit our classical magnetotransport data with the two-component Drude model. 
It allows us to discern holes and electrons at negative gate voltages, and two types of electrons at positive gate voltages. For the latter case the simulated curves (red lines in Fig.~\ref{Fig2}) nearly ideally follow the experimental data.
In comparison to electron side, at negative gate voltages the discrepancy between experiment and fitting is larger. At the optimal set of fitting parameters, the simulated resistance at zero magnetic field is higher than in the experiment, indicating a possible underestimation of the mobility of carriers in this region.
The most likely reason for the poor fit is the  dependence of carrier mobility on the magnetic field, which is not accounted by the model. 
Another possible reason of bad fits is a possible presence of third group of carriers. Next, we will analyze the parameters obtained from the fitting and determine the more likely cause.

\begin{figure}
	\includegraphics[width=1\columnwidth]{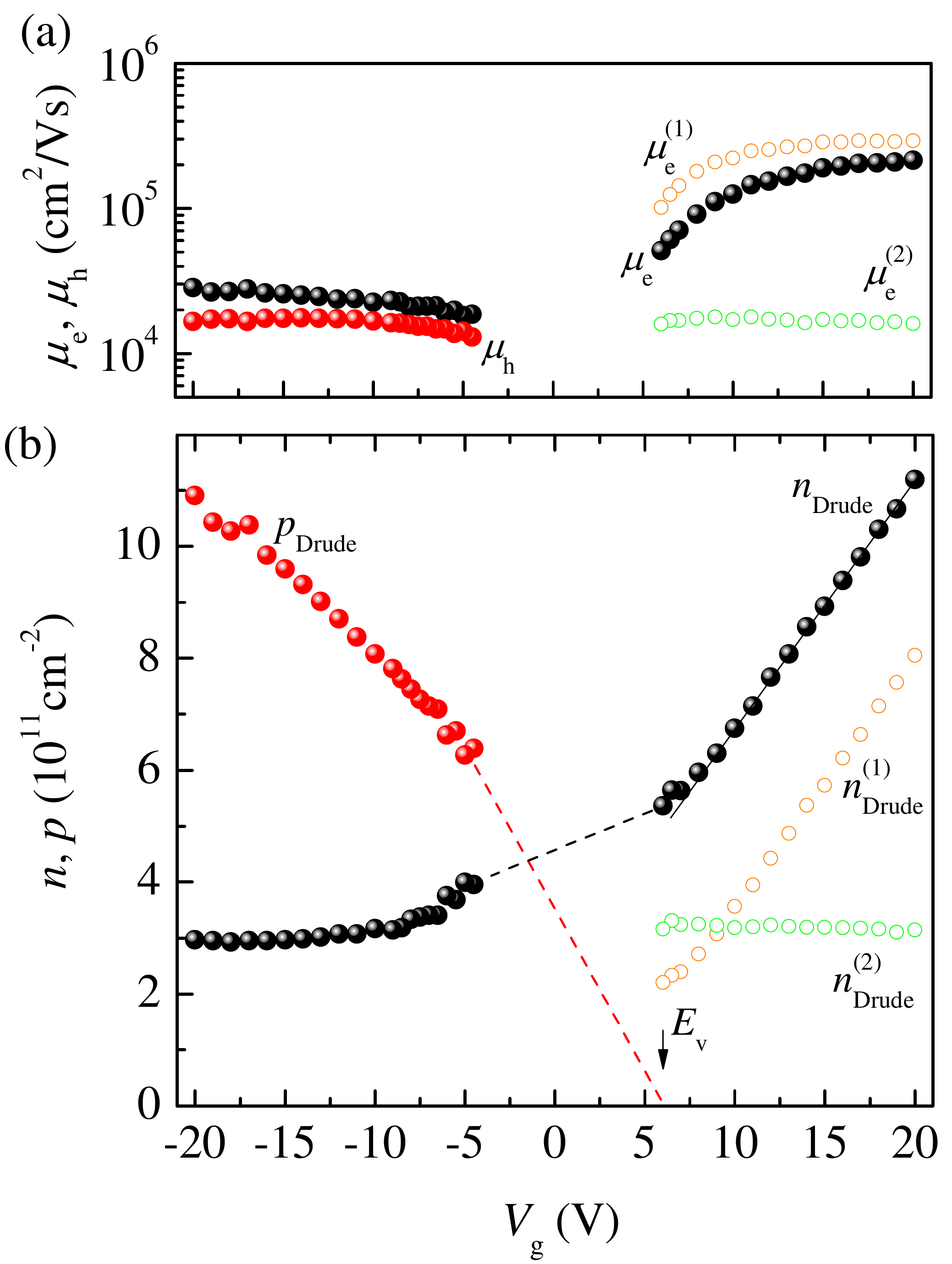}
	\caption{
		The results of the two-component Drude fitting of the classical magnetotransport data.
		(a) The gate voltage dependences of the average electron $\mu_\text{e}$ (black spheres) and hole $\mu_\text{h}$ (red spheres) mobilities, and partial electron mobilities $\mu_\text{e}^{(1)}$ (orange circles) and $\mu_\text{e}^{(2)}$ (green circles). 
		(b) The gate voltage dependences of the total electron $n_\text{Drude}$ (black spheres) and hole $p_\text{Drude}$ (red spheres) densities, and partial electron densities $n_\text{Drude}^{(1)}$ (orange circles) and $n_\text{Drude}^{(2)}$ (green circles).
		A solid black line illustrates the filling rate in the structure $\alpha = 4.4 \times 10^{11}\,$cm$^{-2}$/V, dash black and red lines correspond to the suggested densities behavior, the $E_\text{v}$ position comes from the capacitance measurements in Fig.~\ref{Fig1}\,(c).
	} \label{Fig3}
\end{figure}

Obtained from the Drude fitting gate voltage dependence of electron ($n_\text{Drude}^{(1)}$, $n_\text{Drude}^{(2)}$) 
and hole ($p_\text{Drude}$) densities, as well as their mobilities ($\mu_\text{e}^{(1)}$, $\mu_\text{e}^{(2)}$, $\mu_\text{h}$) 
are shown in Fig.~\ref{Fig3}. 
At negative $V_\text{g}$, where holes (red spheres) and electrons (black spheres) coexist, the performed fitting provides their total densities ($n_\text{Drude}$ and $p_\text{Drude}$) and average mobilities ($\mu_\text{e}$ and $\mu_\text{h}$).
At positive $V_\text{g}$ the first electron density $n_\text{Drude}^{(1)}$ (orange circles) increase nearly linearly with the gate voltage increase, while the second $n_\text{Drude}^{(2)}$ (green circles) is found to be gate independent, so the total electron density $n_\text{Drude} = n_\text{Drude}^{(1)} + n_\text{Drude}^{(2)}$ also linearly increases with $V_\text{g}$ as expected.
The mobility of the first group of electrons $\mu_\text{e}^{(1)}$ (orange circles) is about 10 times higher compare to the mobility of the second group $\mu_\text{e}^{(2)}$ (green circles). 
The maximum value of the averaged electron mobility $\mu_\text{e}=\sum \mu_\text{e}^{(i)} n_\text{e}^{(i)}/(n_\text{e}^{(1)}+n_\text{e}^{(2)})$ is about 3$\times 10^{5}\,$cm$^{2}/$Vs, and the maximum hole mobility values are around $2\times 10^{4}\,$cm$^{2}/$Vs that is consistent with previous studies of thinner 80- and 200-nm HgTe films~\cite{Kozlov2014,Savchenko2019}, indicating the high quality of the growth despite the relaxation of the HgTe lattice.

The total electron density $n_\text{Drude}$ depends linearly on $V_\text{g}$ on the positive-gate-voltage side with the slope of about $4.5\times10^{10}\,$cm$^{-2}$/V that is within the calculated from the electrostatics of the device range ($4.9\times10^{10}\,$cm$^{-2}$/V if the carriers are located at the interface between CdHgTe and HgTe). 
The electron filling rate shows a tendency to decrease at $V_g \lesssim 5 \ldots 7.5$\,V, which is consistent with our valence band top mapping obtained from the capacitance measurements: when electrons and holes coexist, they share the total filling rate proportionally to their densities of states resulting in the decrease of the electron filling rate. 
In the region of $V_g = -5...5$\,V the fitting gives inadequate results because of the vicinity to the charge neutrality point.
At larger negative gate voltages, the fitting works satisfactory again giving qualitatively correct $p_\text{Drude}(V_g)$ dependence, though the slope is less than expected ($d(n_\text{Drude} -p_\text{Drude})/dV_g  = 4\times10^{10}\,$cm$^{-2}$/V for $V_g = -13\ldots-7$\,V).
Keeping all in mind we plot an extrapolated density $p_\text{Drude}$ in the~Fig.~\ref{Fig3}\,(b) with a dashed line with an expected $E_v$ point located at 6\,V.

\subsection{Quantum transport}

\begin{figure}
	\includegraphics[width=0.9\columnwidth]{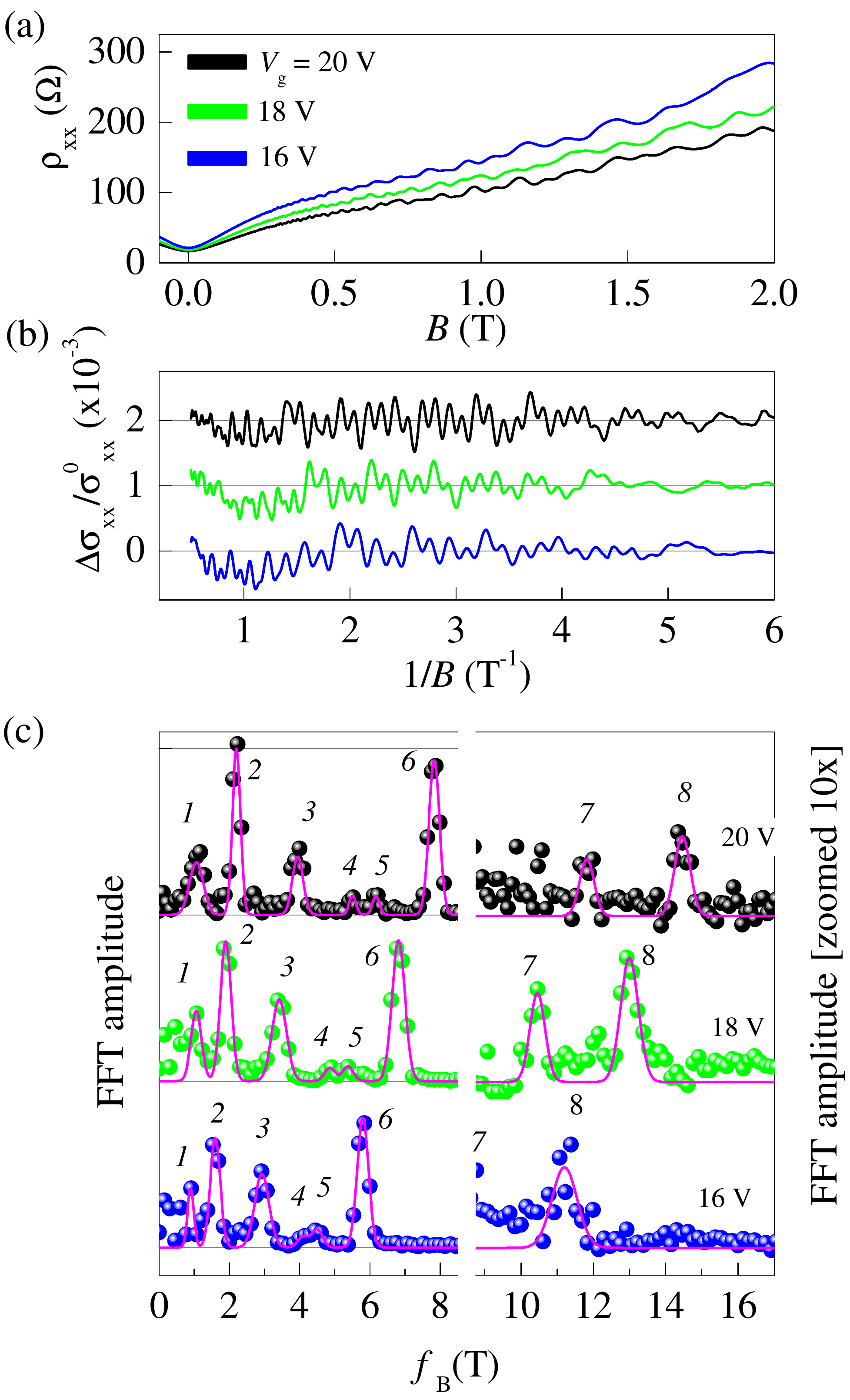}
	\caption{
		(a)~The examples of Shubnikov -- de Haas oscillations measured in $\rho_\text{xx}(B)$ at high positive gate voltages.
		(b)~The corresponding normalized conductivity oscillations $\Delta \sigma_\text{xx}/\sigma_\text{xx}^0 = (\sigma_\text{xx}~ - <\!\sigma_\text{xx} \!>)/\sigma_\text{xx}^0$ in $1/B$ scale, where $<\!\sigma_\text{xx} \!>$ is the monotonous part of conductivity and $\sigma_\text{xx}^0$ is the conductivity at zero field.  
		(c)~The corresponding normalized Fourier spectra of the conductivity oscillations. The solid lines correspond to the fitting by Gaussian functions. The right part of data for $f_\text{B}>8.5$\,T is magnified by 10 times on vertical axis. 		
	} \label{SDH_e}
\end{figure}

The system under study exhibits pronounced Shubnikov -- de Haas (SdH) oscillations, shown in the  
Fig.~\ref{SDH_e} and \ref{SDH_h}. Their analysis  in 3D systems makes it possible to map out the Fermi surface, whereas in 2D structures it is possible to extract directly from the oscillation period the value of carriers density.
Carriers in the studied devices have the De Broglie wavelength that is much lower compare to the thickness of the HgTe film, so they should be considered as a bulk carriers. Due to the zero energy gap and the arbitrary initial distribution of the electrostatic potential, bulk electrons and holes can exist simultaneously.
At the same time, high-mobility 2D carriers are present in the system too, namely: topological surface states at any gate voltage, as well as electrons and holes in the accumulation layer at non-zero gate voltages.
Moreover, magnetotransport measured in parallel magnetic  fields~(see Supplementary Fig.~S1) displays no SdH oscillations, meaning that the observed in perpendicular magnetic fields SdH-oscillations should be treated as coming from the 2D carriers only.

Let's first focus on positive  gate voltages. 
Fig.~\ref{SDH_e}\,(a) demonstrates the examples of the SdH oscillations measured in $\rho_\text{xx}(B)$ at $V_\text{g} = 20$, $18$ and $16$~V. The values of the gate voltages in the figure are deliberately chosen close to each other in order to demonstrate an evolution of the oscillation pattern. 
However, the analysis of oscillations was also carried out in a wider gate voltage range~(see Supplementary Fig.~S3 and S4). 
Each $\rho_\text{xx}(B)$ trace was recalculated in $\sigma_\text{xx}(B^{-1})$, after which its monotonous part $<\!\sigma_\text{xx} \!>$ was subtracted, and the remaining was normalized on the conductivity at zero field $\sigma_\text{xx}^0$. 
The examples of the resulting $\Delta \sigma_\text{xx}/\sigma_\text{xx}^0$ curves are presented in~Fig.~\ref{SDH_e}~(b). The oscillations show a complicated pattern indicating the existence of several groups of 2D carriers. Their Fourier spectra demonstrate 8 peaks, whose position systematically changes with the gate voltage, denoted by frequencies $f^\text{e}_\text{(i=1,\dots,8)}$ and indicated by numbers in the~Fig.~\ref{SDH_e}\,(c). 

\begin{figure}
	\includegraphics[width=1\columnwidth]{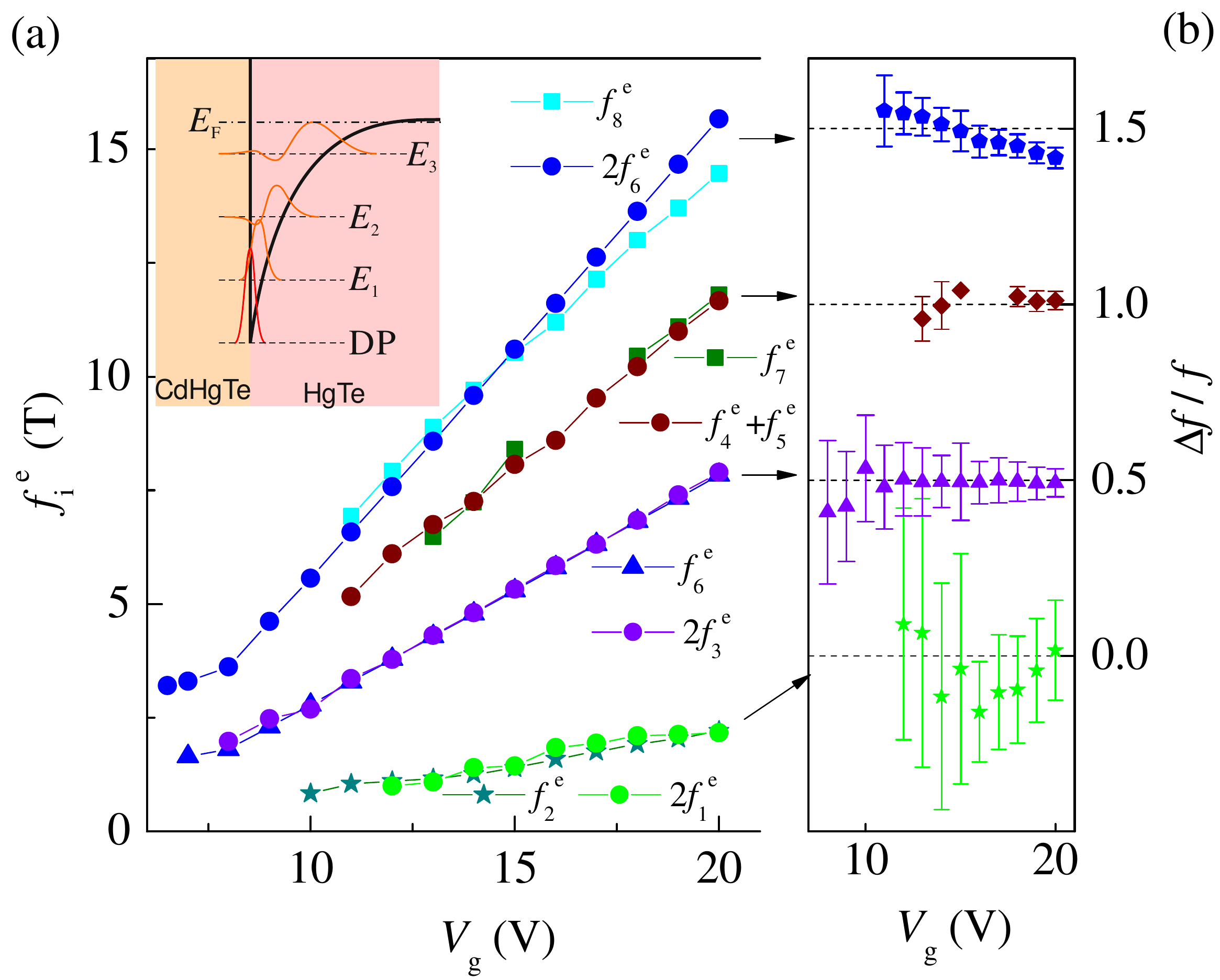}
	\caption{
		(a)~The gate voltage dependence of SdH frequencies and their superpositions. Insert: the suggested band diagram of the accumulation layer for positive gate voltages. $E_{i=1\ldots3}$ denotes edges of electron sub-bands in the accumulation layer, DP is the Dirac point of topological surface electrons, $E_\text{F}$ is the Fermi level.
		(b)~The normalized difference between indicated SdH frequencies (shifted on vertical axis by $0.5$ each for clarity).
		Error bars come from the halfwidths of the corresponding peaks.
	} \label{SDH_correlations}
\end{figure}

\begin{figure}
	\includegraphics[width=1\columnwidth]{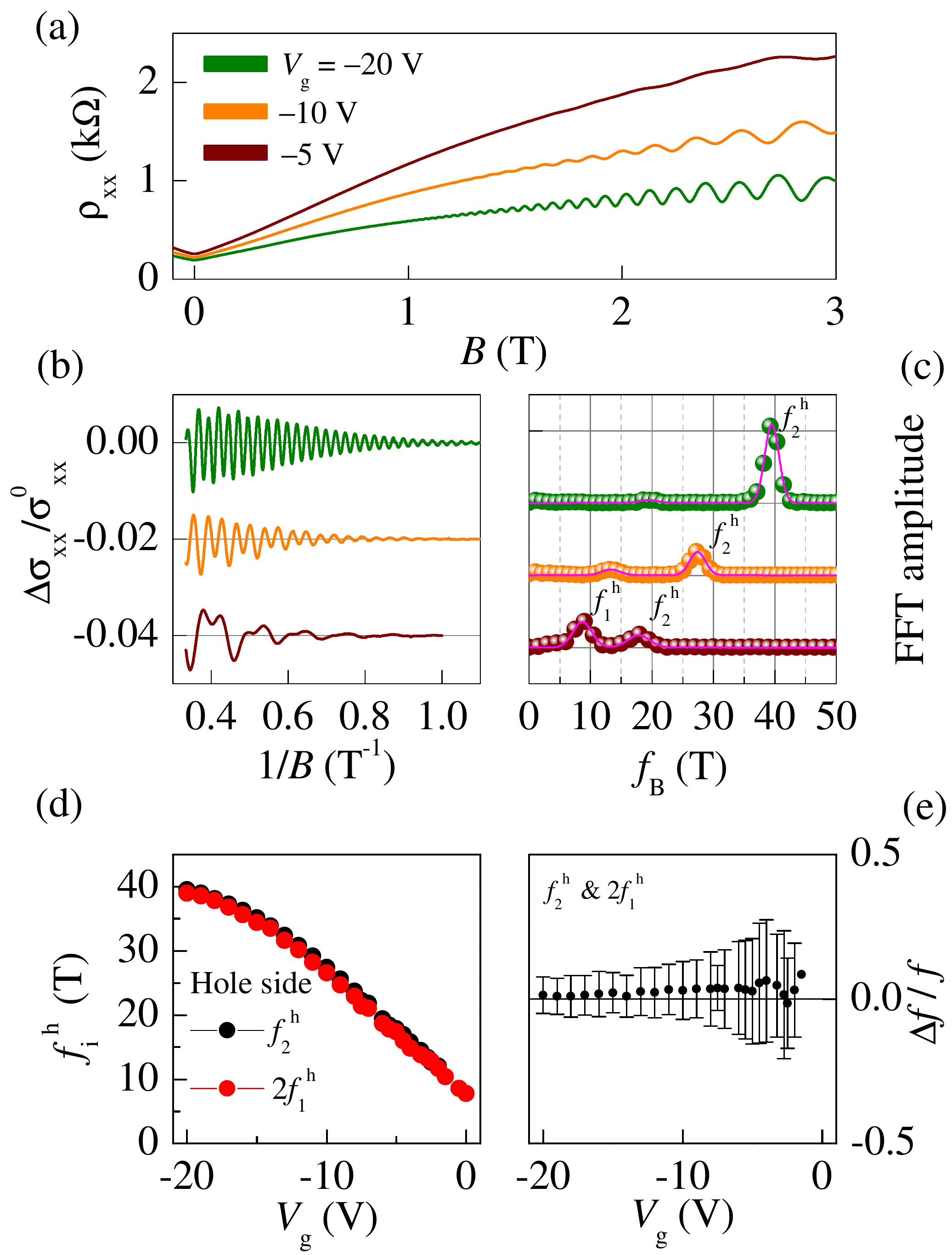}
	\caption{
		(a)~The examples of Shubnikov -- de Haas oscillations measured in $\rho_\text{xx}(B)$ at negative gate voltages.
		(b)~The corresponding normalized conductivity oscillations $\Delta \sigma_\text{xx}/\sigma_\text{xx}^0 = (\sigma_\text{xx}~ - <\!\sigma_\text{xx} \!>)/\sigma_\text{xx}^0$ in $1/B$ scale, where $<\!\sigma_\text{xx} \!>$ is the monotonous part of conductivity and $\sigma_\text{xx}^0$ is the conductivity at zero field.  
		(c)~The corresponding normalized Fourier spectra of the conductivity oscillations.
		The solid lines correspond to the fitting by Gaussian functions. The center of each peak is indicated by the corresponding frequency $f^\text{h}_{i}$.
		(d)~The gate voltage dependence of SdH frequencies $f^\text{h}_{i}$ and their superpositions. 
		(e)~The normalized difference between indicated SdH frequencies $(f^\text{h}_{2} - 2f^\text{h}_{1})/f^\text{h}_{2}$ proves they are the same origin.
		Error bars come from the halfwidths of the corresponding peaks.
	} \label{SDH_h}
\end{figure}

The analysis of such a complex Fourier spectrum must begin with a search for multiple frequencies. 
We tested various combinations of frequencies and found out that the following relationships are fulfilled in the whole range of the gate voltages: $2 f_1^\text{e} \approx f_2^\text{e}$; $2  f_3^\text{e} \approx f_6^\text{e}$; $f_4^\text{e} + f_5^\text{e} \approx f_7^\text{e}$. 
The relationships are shown in~Fig.~\ref{SDH_correlations}\,(a) (three bottom traces) as a $f(V_g)$ dependencies and also as a normalized difference $\Delta f /  f(V_g)$ in~Fig.~\ref{SDH_correlations}\,(b), where $\Delta f$ denotes the difference between left and right sides of relationships  discovered. Thus, the peaks of the Fourier spectrum with numbers from 1 to 7 correspond to three different groups of carriers. 
The first two groups exhibit spin degeneracy of Landau levels at small magnetic fields, which is lifted as magnetic field increases resulting in a simply doubling of the oscillation frequency. 
In contrast, the third groups of carriers ($f_4^\text{e}$ and $f_5^\text{e}$) shows different behaviour, typical for Rashba-splitted electrons~\cite{Novik2005, Khudaiberdiev2022}: at small magnetic fields it shows a beating oscillation pattern which is reflected by two closely spaced Fourier peaks with a transition to a separated Landau levels at higher fields with the frequency $f_7^\text{e} = f_4^\text{e} + f_5^\text{e}$. 
We also found that $2 f_6^\text{e} \neq f_8^\text{e}$ (upper traces in the~Fig.~\ref{SDH_correlations}). The eighth Fourier peak has no harmonics of either higher or lower frequency and therefore corresponds to a fourth group of carriers without spin degeneracy, i.e., topological surface electrons. One should note that the amplitude of this peak is too small for carriers of a such high density and respective contribution to the total conductivity. Although the nature of this phenomenon is not clear, the trend was already observed during the comparison of 80\,nm HgTe films, where the topological electrons on the top surface made the main contribution to the SdH oscillations~\cite{Kozlov2014, Ziegler2020}, with 200\,nm, where their contribution was already several times weaker compared to other carriers~\cite{Savchenko2019}. Extrapolating this trend to the fully relaxed 1000\,nm film under study, one should expect that the peak in the Fourier spectrum from the topological electrons might be significantly damped.
Thus, we believe that electron SdH  oscillations are formed by four groups of carriers, three of which have spin degeneracy and one does not. 

The existence of 2D carriers with spin degeneracy is not surprising. 
They can be both bulk electrons in the accumulation layer formed at positive gate voltages (see insert in the Fig.~\ref{SDH_e}) or Volkov-Pankratov states~\cite{Volkov1985}~(VPS). The latter have the same origin as the topological surface states~\cite{Inhofer2017, Tchoumakov2017}, but they form only if there is a smooth transition between topologically trivial and non-trivial materials, or if there is strong enough band bending near the boundary of materials with the opposite topology~\cite{Tchoumakov2017}. 
We expect to have a sharp interface between HgTe and CdHgTe barriers since it can be controlled well during the epitaxy growth~\cite{Otteneder2020}, and suggest the induced by the gate band bending creates conditions for the VPS  formation. For our study, the difference between VPS and bulk electrons in the accumulation layer is immaterial. Therefore, for convenience and to avoid confusion with bulk electrons located far from the gate outside the accumulation layer, we will refer to all three detected groups of spin-degenerated carriers as accumulation layer electrons. 

\begin{figure*}
	\includegraphics[width=2\columnwidth]{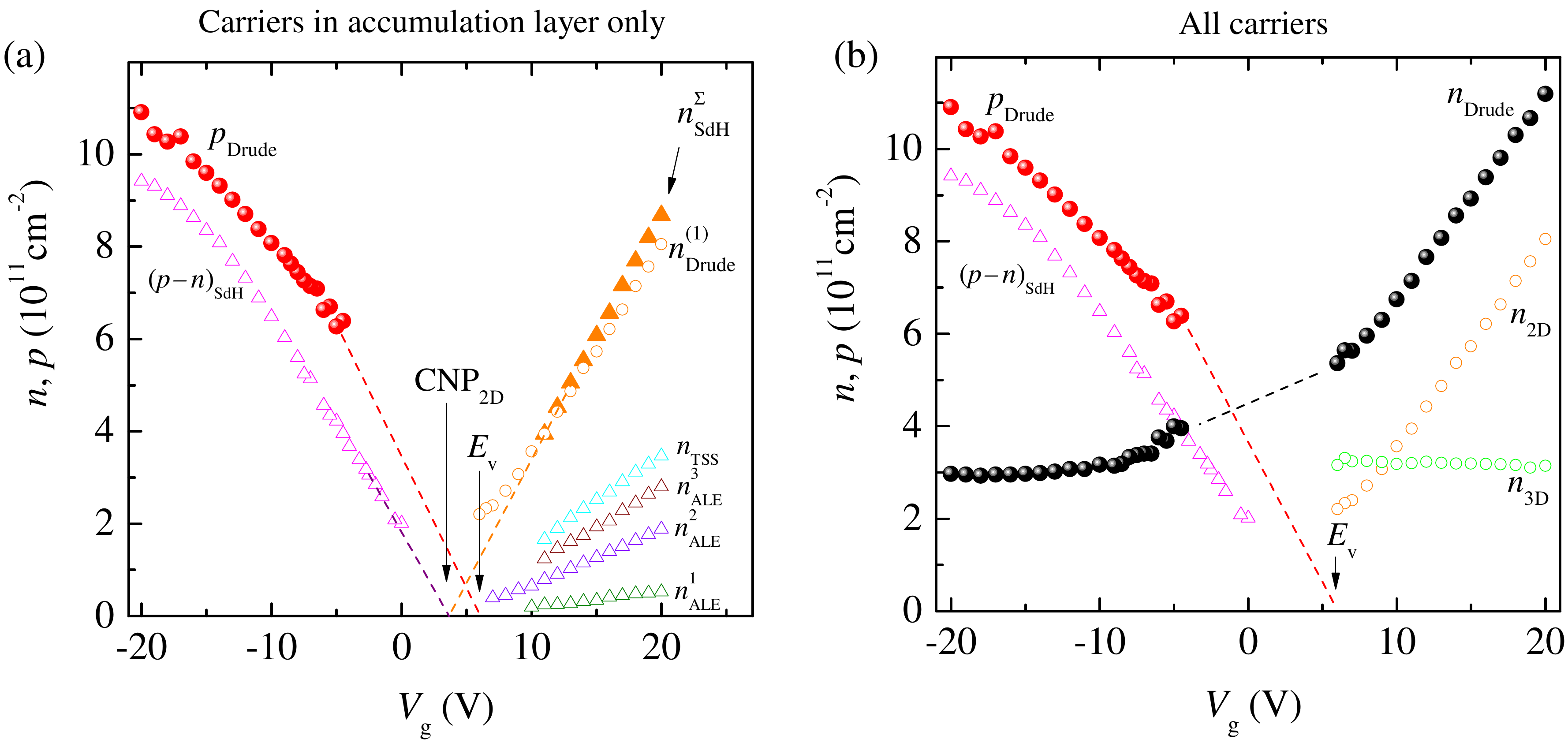}
	\caption{
 	(a)~The comparison of electron and hole densities obtained from SdH oscillations  and from the Drude fitting (data from Fig.~\ref{Fig3}\,(b)). The densities obtained from the Drude fitting shown with red spheres ($p_\text{Drude}$) and  orange circles ($n_\text{Drude}^{(1)}$). The SdH oscillations period reflects the densities of 2D carriers located in the vicinity of the gate, namely in the accumulation layer and topological surface states, shown by triangles. In the valence band it reflects the diffrential denisty $(p-n)_\text{SdH}$ (magenta), while on the electron side we are able distinguish three types of the accumulation layer electrons ($n_\text{ALE}^{1\ldots3}$,  olive, violet and brown, the color matches with the ones in the Fig.~\ref{SDH_correlations}) and topological surface electrons ($n_\text{TSS}$, cyan). The total 2D electron density $n_\text{SdH}^\Sigma = \sum n_\text{ALE}^{1\ldots3} + n_\text{TSS}$ satisfactorily matches to the high-mobility electron density $n_\text{Drude}^{(1)}$.
	(b)~The combined gate voltage -- density map,
	$n_\text{Drude}$ and $p_\text{Drude}$ have the same color code as in Fig.~\ref{Fig3}\,(b) and are shown as black and red spheres, respectively.
	Orange circles represent 2D electrons of density $n_\text{2D}$, green circles -- bulk 3D electrons of density $n_\text{3D}$ (see text for details).	
	} \label{Fig7}
\end{figure*}

The accumulation layer electrons (ALE) together with topological surface states (TSS) give a self-consistent picture. 
First, we introduce their partial densities as $n^\text{k=1\ldots 3}_\text{ALE}$ and $n_\text{TSS}$ accordingly. The electron density of each group can be determined by the formula $g_s^i \frac{e}{h} f_i^\text{e}$, where $i=1\ldots8$ is the peak index, $g_s^i$ is the appropriate spin degeneracy, $h$ is the Planck constant: $n^\text{1}_\text{ALE} = 2 \frac{e}{h} f_1^\text{e} = \frac{e}{h} f_2^\text{e}$, $n^\text{2}_\text{ALE} = 2 \frac{e}{h} f_3^\text{e} = \frac{e}{h} f_6^\text{e}$, $n^\text{3}_\text{ALE} =  \frac{e}{h} f_4^\text{e} + \frac{e}{h} f_5^\text{e} = \frac{e}{h} f_7^\text{e}$ and $n_\text{TSS} = \frac{e}{h} f_8^\text{e}$. The density dependencies on the gate voltage obtained in this manner are shown in Fig.~\ref{Fig7}(a). The total density of 2D carriers obtained $n_\text{SdH}^\Sigma$ agrees well with the density $n_\text{Drude}^\text{(1)}$ of high-mobility electrons determined from the two-component Drude model fitting for positive gate voltages. Second, carriers with a higher density also have a higher filling rate, which indicates their closer location to the gate and in line with the model of  the triangular potential accumulation layer. Third, only one group of accumulation layer electrons with the highest density exhibits Rashba splitting. This fact is consistent with the assumption that this group of electrons is closest to the gate (with the exception of topological electrons, for whom the Rashba splitting is irrelevant), where the electric field is strongest. 

To sum up the electron side, we found that four group of carrier (three groups of accumulation layer electrons and one topological) are contributing to SdH oscillations and their total density matches with the density of high-mobility carriers obtained from the two-component Drude fitting. The low-mobility electrons, apparently located in the bulk and not affected by the gate, acts as a background. 
We also do not see any traces of back surface topological electrons. This seems to be due to their low density.

Now we switch to the negative gate voltages. The obtained 
$\rho_\text{xx}(B)$ traces for $V_\text{g} = -20, -10$ and $-5$\,V are shown in the~Fig.~\ref{SDH_h}\,(a). Following the same procedure as for the electron side we extracted the oscillatory part of the conductivity, shown in the~Fig.~\ref{SDH_h}\,(b). The Fourier spectra of oscillations are shown in the~Fig.~\ref{SDH_h}\,(c). 
Two distinct peaks can be identified, marked by $f^\text{h}_{1}$ and $f^\text{h}_2$, that correspond to the formation of spin-degenerate and resolved Landau levels, respectively. The spin-degenerate frequency $f^\text{h}_{1}$ is seen better at closer to zero $V_\text{g}$ and lower $B$, while the $f^\text{h}_{2}$ frequency is more pronounced under opposite conditions 
reflecting the change in the relation between Zeeman and orbital splitting.
The gate voltage dependences of $f_2^\text{h}$ and $2f_1^\text{h}$ are shown in Fig.~\ref{SDH_h}\,(d), 
where it is clearly seen that the ratio $f^\text{h}_2/f^\text{h}_1 = 2$ holds. Additionally, the normalized difference between indicated SdH frequencies $(f^\text{h}_{2} - 2f^\text{h}_{1})/f^\text{h}_{2}$ also shows almost zero value in the~Fig.~\ref{SDH_h}\,(e) proving that these peaks have the same origin.

In single-component 2D systems, the SdH oscillations frequency reflects the charge carrier density of electrons or holes. 
In multi-component systems, where holes coexist with electrons, like in HgTe films with a thickness of 80-200\,nm~\cite{Mendez1985, Raichev2012b, Kozlov2014, Savchenko2019}, the period of oscillations in the valence band reflects the  differential density of holes and electrons, i.e.,  $(p-n)_\text{SdH} = 2(e/h) f_1^\text{h}$. The obtained by this manner dependence  
$(p-n)_\text{SdH}(V_g)$ is shown in the Fig.~\ref{Fig7}~(a). It is clearly seen that $(p-n)_\text{SdH}$ shows systematically lower values than $p_\text{Drude}$. One could extrapolate both $(p-n)_\text{SdH}(V_g)$ and $n_\text{SdH}^\Sigma(V_g)$ dependencies to zero and found that they cross the horizontal axis at the same point, namely at 3\,V. Apparently, this is the charge neutrality point 
CNP$_\text{2D}$ for all 2D carriers (both electrons and holes located in the accumulation layer and topological electrons). The position of this point is consistent with the valence band top position, located on the right side from the CNP$_\text{2D}$, at 5$\ldots$7.5\,V. 

Fig.~\ref{Fig7}\,(b) summarizes our findings. Here we show obtained from the Drude fitting hole $p_\text{Drude}$ and total electron $n_\text{Drude}$ densities, as well as partial 2D $n_\text{2D} = n_\text{Drude}^{(1)}$ and 3D $n_\text{3D} = n_\text{Drude}^{(2)}$ densities, for electrons located near the gate and in the bulk accordingly.
At high positive $V_\text{g}$ there is a mixture of 3D and 2D electrons in the system. The latter ones are high-mobility electrons that consist from topological surface electrons, Volkov-Pankratov and trivial electrons at the accumulation layer.
The change of the gate voltage results in the change of the profile of the electrostatic potential of the accumulation layer, while the Fermi level of HgTe is pinned by the 3D bulk electrons of constant density $n_\text{3D}$.
At lower gate voltages, 
at $V_\text{g} = 5\ldots7.5\,$V, we start to introduce 2D holes in the accumulation layer. From our data it is not clear, if electrons and holes in the accumulation layer co-exist at $V_g < 5$\,V, however it is possible (for instance, co-existence of 2D holes and topological electrons). On the other hand, near the CNP$_\text{2D}$ the accumulation layer is not well-developed yet, while deeper in the valence band ($V_g<0$\,V) we do not see any manifestation of 2D electrons in SdH oscillations. This suggests that their density and/or quantum mobility are too low.  

\subsection{Peculiarities of a transport response}\label{SubSec: Peculiarities}

Here we want to shortly stress out the peculiarities of the quantum transport response of the studied 1000-nm HgTe films.
The first feature is an already discussed rather weak oscillatory response of the topological surface states compare to other types of electrons. 
This is especially strange since they have the highest density and, presumably, high mobility.
Perhaps, their proximity to the scattering centers (the CdHgTe/HgTe interface) plays a role here, 
and the topological protection against backscattering~\cite{Hasan2010, Ando2013}, which may increase the transport scattering time $\tau_{tr}$, has a little or no effect on the quantum time $\tau_q$, responsible for the SdH oscillations amplitude.

\begin{figure}
	\includegraphics[width=1\columnwidth]{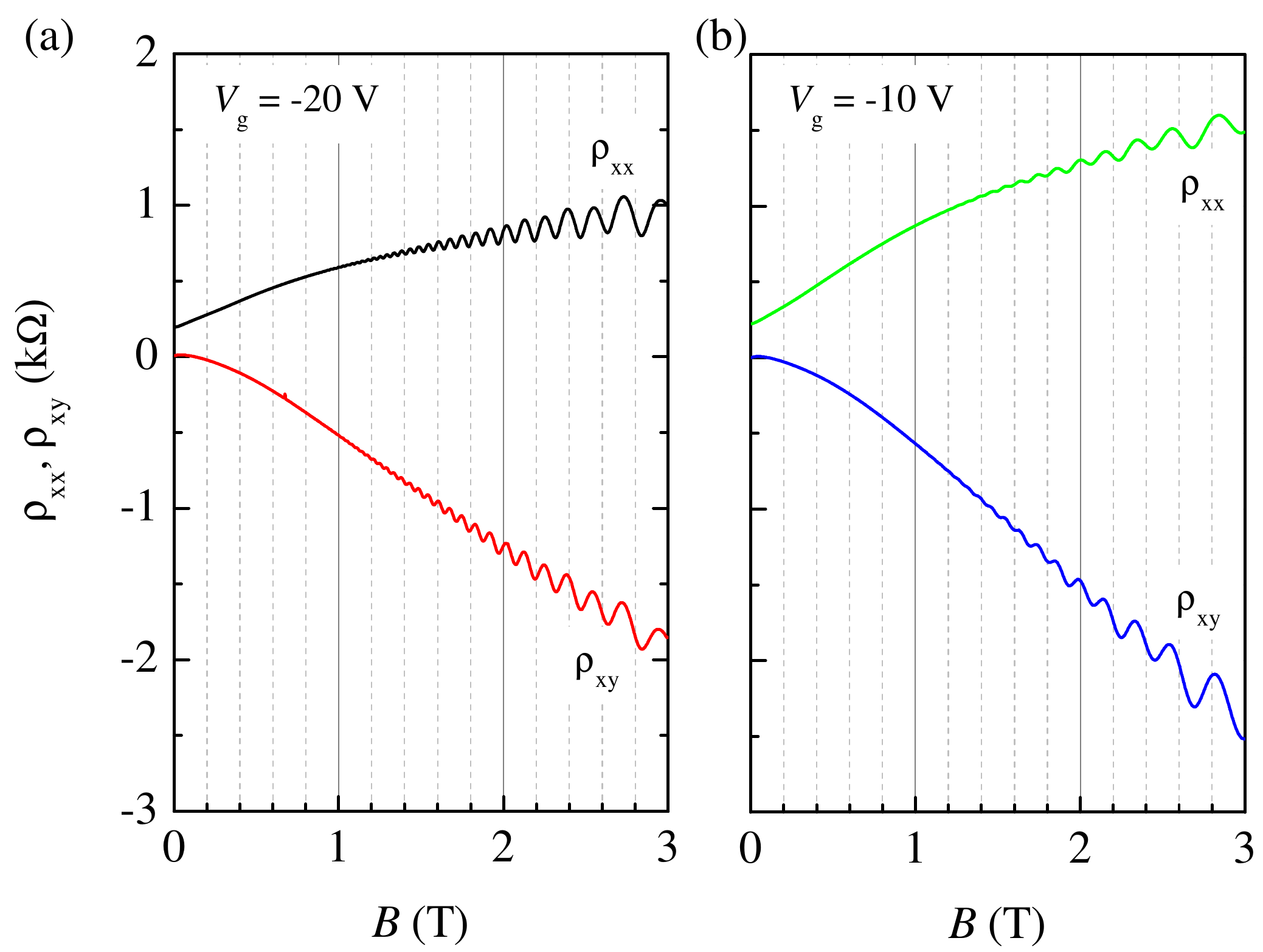}
	\caption{
		The examples of Shubnikov -- de Haas oscillations measured in $\rho_\text{xx}(B)$ and $\rho_\text{xy}(B)$ at $V_\text{g} = -20$~(a) and $-10$~V~(b).
	} \label{Fig9}
\end{figure}
The second feature for discussion is anomalously strong SdH oscillations observed in the Hall resistance at negative gate voltages~(Fig.~\ref{Fig9}). In comparison to oscillations in $\rho_\text{xx}$, they are characterized by the same period and nearly the same amplitude while having an opposite phase  ~\cite{Isihara1986, Coleridge1989, Mani2009, Minkov2023}.
In general, the SdH oscillations of both diagonal and  Hall components of the resistivity tensor stem from the oscillatory density of states~\cite{Ihn2010}.
However, the oscillating part of  
$\rho_\text{xy}$ survives only in the case of short-range scattering~\cite{Isihara1986, Ihn2010}, that we expect as the main scattering mechanism in the accumulation layer. Next, according to the theories~\cite{Isihara1986, Coleridge1989, Minkov2023}, the amplitude of oscillations in $\rho_\text{xy}$ is additionally damped in comparison to the one in $\rho_\text{xx}$ with a damping factor of $1/\mu B$.
In our case $1/\mu B \ll 1$, however the observed oscillation amplitude is nearly the same for both $\rho_\text{xx}$ and $\rho_\text{xy}$, which is anomalous.
Some differences between experiment and theory for the SdH amplitude in thinner, $(5-30)\,$nm-thick, HgTe systems was also observed in~\cite{Minkov2023}, but apart from our findings, there the main ratio $\Delta \rho_\text{xx} \gg \Delta \rho_\text{xy}$ holds.
We also note that, despite thinner, 20-, 80-, and 200-nm HgTe quantum wells host electrons and holes of similar properties, we did not observed such pronounced SdH oscillations of $\rho_\text{xy}$ there.

\section{Conclusion}
In summary, we have shown that both 2D and 3D carriers may present, depending on the Fermi level position in 1000-nm HgTe film. 
The 2D carriers are located on the interface between HgTe and CdHgTe (topological electrons) or in the vicinity of it (trivial electrons or holes, or Volkov-Pankratov electrons), in formed by the gate voltage
accumulation layer. Both 2D electrons and holes exhibit pronounced Shubnikov-de Haas oscillations, sensitive to the perpendicular component of the magnetic field. 3D electrons are located in the bulk, act as separate classical conductive channel and pins the Fermi level, making this system an ideal candidate to further study of the Quantum Hall effect reservoir model~\cite{Zawadzki2014, Tyurin2010}.

\section{Acknowledgements}
The work is supported by RFBR Grant No. 18-32-00138. 

\bibliography{library}

\pagebreak
\widetext
\begin{center}
	\textbf{SUPPLEMENTAL MATERIAL}
\end{center}

\setcounter{figure}{0}
\setcounter{equation}{0}
\setcounter{section}{0}
\renewcommand{\thesection}{S\arabic{section}}
\renewcommand{\theequation} {S\arabic{equation}}
\renewcommand{\thefigure} {S\arabic{figure}}
\renewcommand{\thetable} {S\arabic{table}}

%
%
%
%
%
%
%
%
%
%
%
	
\section{Parallel fields}\label{append: Parallel fields}

\begin{figure}[h]
	\includegraphics[width=1\columnwidth]{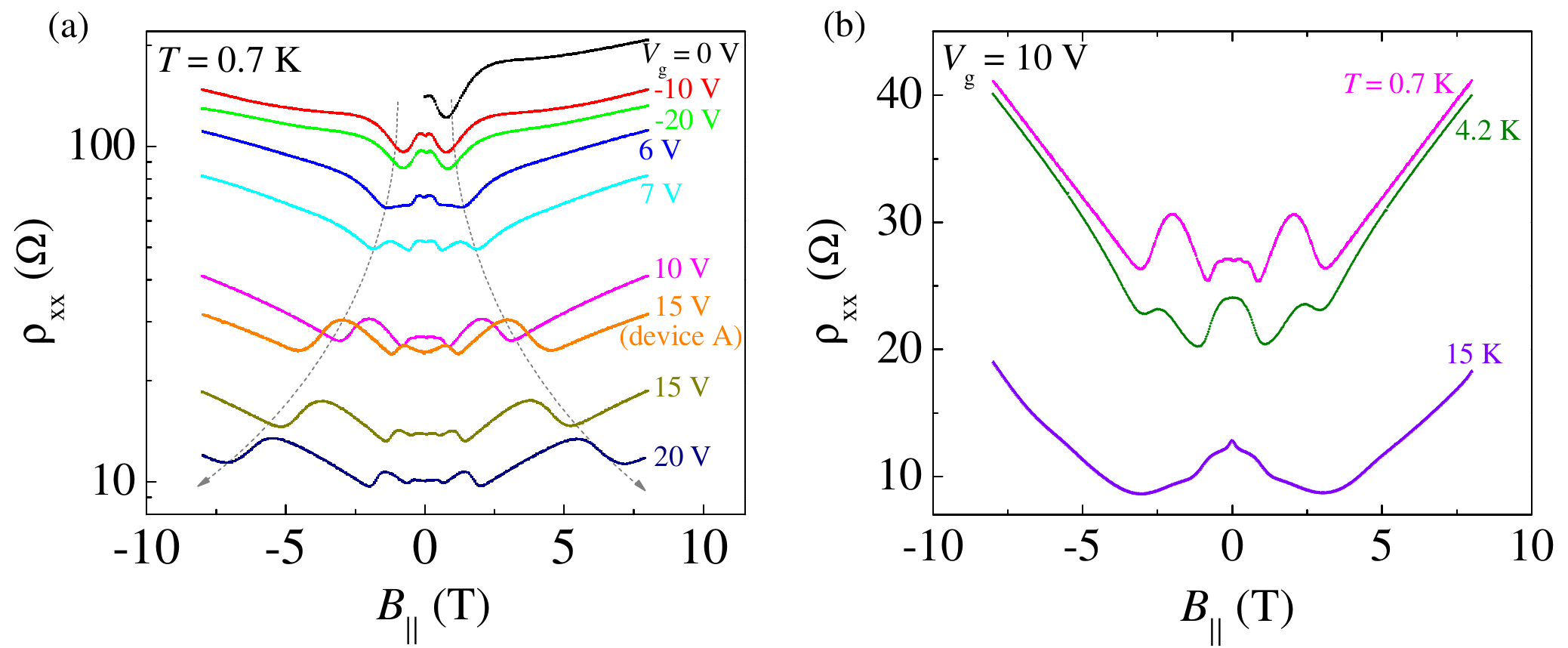}
	\caption{
		(a)~The diagonal resistance vs parallel magnetic field measured at different gate voltages and 
		at fixed temperature 0.7\,K. 
		Orange curve at $V_\text{g} = 15\,$V (``device A'') show the data obtained on the device discussed in the main text.
		It is seen that the observed in the $\rho_\text{xx}(B_{||})$ dependence features do not scale with the density but rather reflect the band structure modification. 
		(b)~The diagonal resistance vs parallel magnetic field measured at different temperatures and 
		at the fixed gate voltage 10\,V. 
	} \label{FigS1}
\end{figure}

Despite nominally the studied 1000-nm HgTe films host 3D carriers, any nonzero gate voltage forms an accumulation layer that forms a 2D subsystem. 
The existence of these 2D carriers results in the observation of shown in the main text SdH oscillations measured at perpendicular magnetic field.
In Fig.~\ref{FigS1} we also show the dependence of diagonal resistance on the parallel magnetic field $B_{||}$ measured at different gate voltages~(a) and temperatures~(b). 
It is seen that the observed on $\rho_\text{xx}(B_{||})$ features do not scale with the density but reflect the band structure modification. 

\newpage
\section{All $\rho_\text{xx}$ and $\rho_\text{xy}$ vs $B$ data}\label{append: R(B)}

\begin{figure}[h]
	\includegraphics[width=0.9\columnwidth]{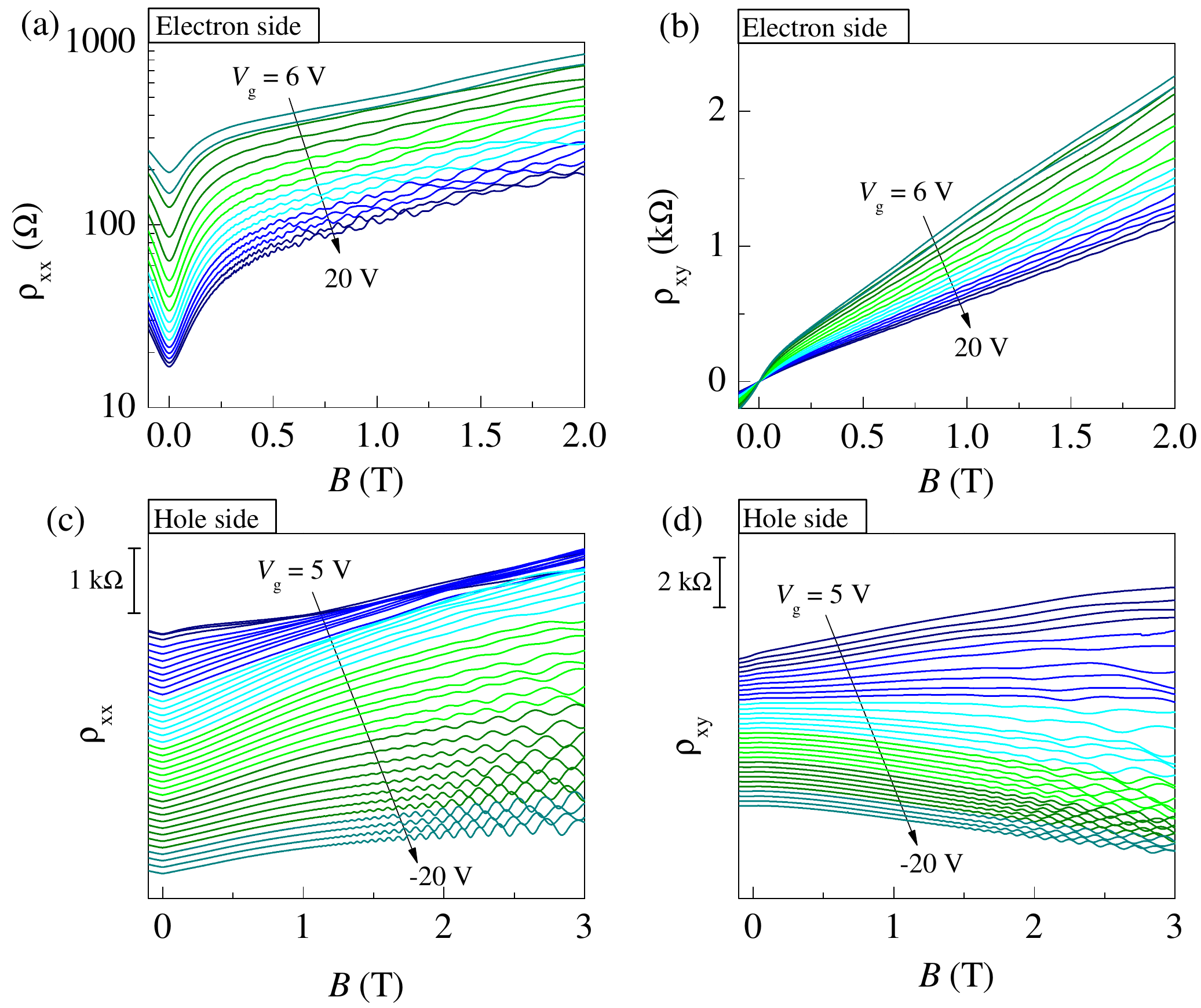}
	\caption{
		Perpendicular magnetic field dependences of $\rho_\text{xx}$~(a), (c), and $\rho_\text{xy}$~(b), (d) measured at different gate voltages. 
		Curves on panels~(c) and (d) are shifted for clarity.
	} \label{FigS2}
\end{figure}

\newpage
\section{SdH oscillations at all gate voltages}\label{append: FFT}

\begin{figure}[h]
	\includegraphics[width=1\columnwidth]{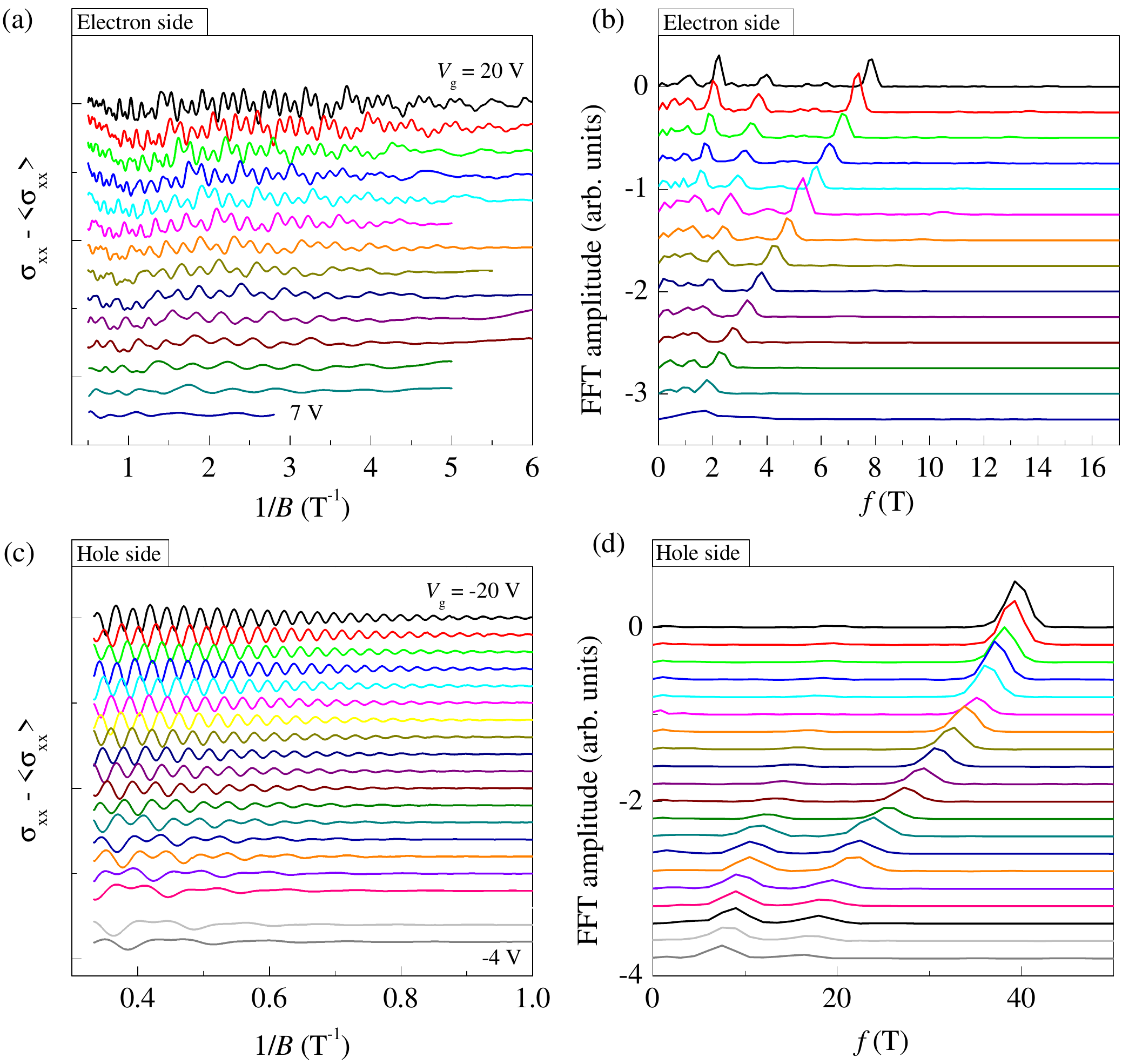}
	\caption{
		(a)~and (c) Shubnikov -- de Haas oscillations of conductivity $\Delta \sigma_\text{xx} = (\sigma_\text{xx}~ - <\!\sigma_\text{xx} \!>)$ measured at positive and negative gate voltages, respectively.
		(b)~and (d) The corresponding Fourier spectra of the conductivity oscillations.
	} \label{FigS3}
\end{figure}

\newpage
\section{SdH oscillations close to zero gate voltages}\label{append: FFT_CNP}
\begin{figure}[h]
	\includegraphics[width=0.7\columnwidth]{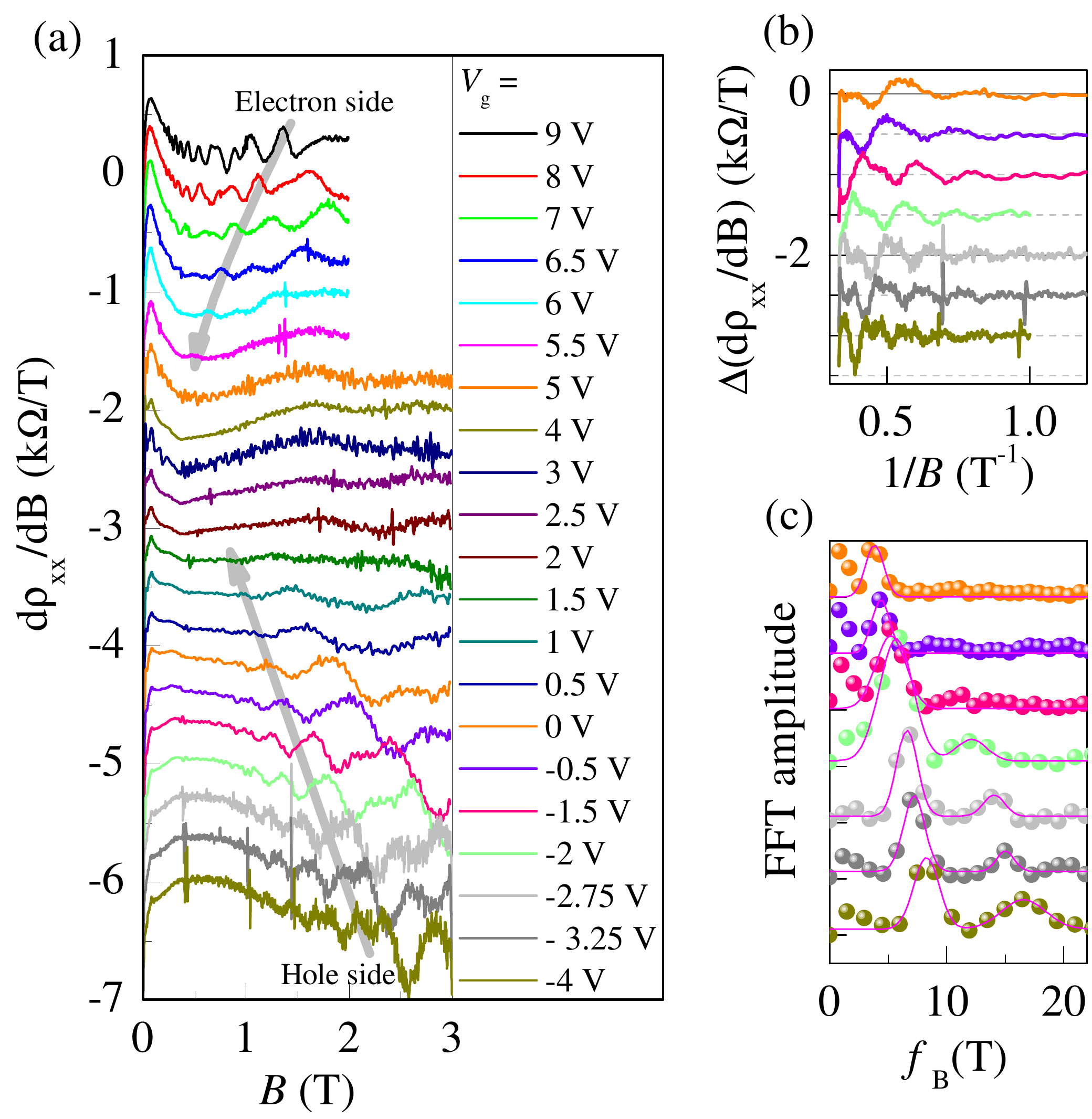}
	\caption{
		(a)~The modification of the Shubnikov -- de Haas oscillations in $\text{d}\rho_\text{xx} / \text{dB}$ measured at different gate voltages moving from the electron side to the hole side.
		(b)~The oscillations in $\Delta (\text{d}\rho_\text{xx} / \text{dB}) = \text{d}\rho_\text{xx} / \text{dB}~ - <\! \text{d}\rho_\text{xx} / \text{dB} \!>$ in $1/B$ scale in the CNP region at $V_\text{g} = 0, -0.5, -1.5, -2\,$V, where $<\! \text{d}\rho_\text{xx} \!>$ is the monotonous part of  $\text{d}\rho_\text{xx} / \text{dB}$.  
		(c)~The corresponding normalized Fourier spectra of the oscillations.
		The solid lines correspond to the fitting by Gaussian functions.
	} \label{Fig8}
\end{figure}
%

\end{document}